**« *Sound people* speak to *Star people* – A sound experts' perspective on astronomy sonification projects »**

N. Misdariis[a], E. Özcan[b], M. Grassi[c], S. Pauletto[d], S. Barrass[e], R. Bresin[d], P. Susini[a]

[a] STMS Ircam-Cnrs-SU, Paris, France
[b] Faculty of Industrial Design Engineering, Delft Univ. of Technology, Delft, The Netherlands
[c] Department of General Psychology, University of Padova, Italy
[d] Dept of Media Technology and Interaction Design – School of Electrical Engineering and Computer Science – KTH Royal Institute of Technology, Stockholm, Sweden
[e] sonification.com , Canberra, Australia

## Abstract

The Audible Universe project aims at making dialogue between two scientific domains investigating two distinct research objects, briefly said, Stars and Sound. It has been instantiated within a collaborative workshop that started to mutually acculturate both communities, by sharing and transmitting respective knowledge, skills and practices. One main outcome of this exchange was a global view on the astronomical data sonification paradigm that allowed to observe either the diversity of tools, uses and users (including visually-impaired people), but also the current limitations and potential ways of improvement. From this perspective, the current paper presents basic elements gathered and contextualised by sound experts in their respective fields (sound perception / cognition, sound design, psychoacoustics, experimental psychology), in order to anchor sonification for astronomy in a more well-informed, methodological and creative process.

## Introduction

Despite the fact that we are all basically blind to the Universe, astronomers have nearly always opted to represent the data collected with visual images. However, taking advantage of the capacity of audition to complement or supplement vision, different sonification approaches have been recently built for different goals (e.g., education or research) and audiences (e.g., astronomers or amateurs, Blind-Visually Impaired (BVI) or sighted persons). In this attempt to dialogue between two specific scientific fields – astronomy and sound – it is desirable to come up with shared knowledge, co-constructed ideas, relevant tools and, at the end, evaluation guidelines that should be as general, comprehensive and efficient as possible, although potentially dependent on the nature of the goals and/or the audiences.

The present paper, written by sound design/perception experts, reports on what they saw and understood of current astronomy sonification projects during an interdisciplinary workshop. It aims to strengthen this interdisciplinary dialogue by providing practical advices on how the astronomy community could draw upon the expertise of the sound community to make progress in the approach of listening to the Universe, instead of, or in addition to, just watching it.

## 1. Context of the Perspective

### 1.1 A participatory workshop as research framework

The 'Audible Universe' aims to establish a collaborative framework to share and develop knowledge, ideas and applications concerning sonification in astronomy. It was initiated by a (remote) workshop in 2021 where nearly 50 experts from different scientific disciplines related to astronomy and sound met and worked together[1]. During the workshop 'Star people' and 'Sound people' shared their respective expertise in an acculturation process of collaborative evaluation and design. The process provided a basis for further developing data sonification as a technique in the handling of astronomical data – whilst also addressing the accessibility issue for the BVI community which is another key concern of the 'Audible Universe' framework.

The experts in astronomy broadly described the specific nature of astronomical data (light curve, spectrum, image, time series, etc.), together with some of the main astronomy-focused sonification tools that are already currently used by researchers in that domain[2]. Detailed information about five of these tools was presented:

• AstreOs (https://astreos.space): a stargazing multi-sensorial astronomy application based on a standardized visualizer in astronomy – Aladin (https://aladin.u-strasbg.fr/) – associated with a sound synthesis engine that mainly maps the brightness value of the RGB components of an image to pure tone audio and haptic clicks, with an additional spatial rendering functionality.

• Starsound (https://www.jeffreyhannam.com/starsound): a standalone application that generically maps any kind of astronomical data to the basic audio dimensions (frequency, intensity, duration) of different types of sound (pure tones, pulses, MIDI (Musical Instrument Digital Interface) instruments).

• SonoUno (http://sion.frm.utn.edu.ar/sonoUno/): another generic application developed methodologically in a designerly way (functionality and ergonomics), based on basic audio parameters (frequency, amplitude and instrument type) ; with an additional screen reader that uses the speech based auditory user interface technique[3].

• A4BD (https://www.a4bd.eu/fr/): a didactic application that teaches the use of haptic vibrations to detect different kinds of edges and shapes (square, triangle, circle, etc.), and audio to detect the colors of an image (hues map to pitches / musical notes and lightness to loudness).

• Afterglow access (https://afterglow.skynetjuniorscholars.org/en/): an online application associated with a sky viewer tool (Skynet Robotic Telescope Network) with the main goal to identify and locate targets of interest (e.g., a star in an image) or observe other astronomical information (eg. track saturation). The application is built on the model of a reading head that parses the 2D-image as a two-dimensional mapping to audio (time and frequency).

Then, in turn, sound experts described the fundamentals of sound design, sonification, sound perception and psychoacoustics. This basic sonic knowledge provides a shared 'toolkit' for analysis, creation and validation in further collaborative works. Three structural questions, formally based on a conventional 3-step design / sound design process[4], were used to motivate discussions between Sound people and Star people:
1. What can we learn from the existing tools, in terms of sound perception and, more largely, sound experience?
2. Where could we be heading, in terms of designing improved or even new sonification tools?
3. How can we evaluate the usefulness, usability and even desirability[5,6] of existing tools?

### 1.2 Outcomes of the first workshop edition

This paper presents the issues that were raised during the Audible Universe workshop, and pointed out during several Question and Answer (Q&A) sessions and discussions[7]. The perspective of the sound experts on sonification in astronomy applications is presented as an action grid that summarises the issues and outcomes raised during the live plenary Q&A sessions and later collective discussions.

In summary, the various following key notions, and questions, appear to be definitely noteworthy.

*Universality.* Could sonification be inserted into a universal design paradigm by taking into account nature and diversity of its audience, but also by considering feasible solutions for both visually-impaired and sighted people?

*Standardization.* Could sonification standards, or guidelines, for astronomical data be required, and in any case, how could it be compatible with a certain level of customization able to ensure adaptability?

*Skepticism.* What would be the right way to overcome a latent skepticism: better understand the uses and users, propose evidence-based design, raise awareness on the added-values of sound (e.g., by a gamification approach)?

*Multimodality.* How could we complement the sound medium with other rendering modalities such as spatiality or haptics, or even through the tangibility of a 3D-printed mediation object[8] ?

*Analogy.* Could we consider the Universe as a complex sound scene, and therefore transpose Auditory Scene Analysis paradigms such as grouping/segregation[9], or acoustic ecology concepts such as acoustic niches[10,11] (spectro and/or temporal zones in sound spectrum where acoustic energy is preferably located)?

*Prototypicality.* Could (a certain form of) sonification be seen at a 'quick and dirty process' for auditorily monitoring astronomical data, as visual representations could operate in some way?

On another side, it is worth noting that some points relevant to fundamental of design were barely (or not at all) discussed but might however stay of major interest for future thoughts and works (they may be left aside due to lack of time, understanding, reference in the domain, …). These include emotions, multiculturality, interactivity, training, and importantly artificial intelligence.

## 2. From Sound Perception to Sound Experience

### 2.1 Basics of sound perception and cognition

Although we can perceive sounds when dreaming or via direct stimulation of our brain, in the majority of cases sound sensation begins when a physical sound wave sets into vibration the eardrum. The human sensation of sounds unfolds along three dimensions (plus one).

The first dimension is *pitch* that is related to the sound's frequency so that sounds can be perceived as low or high in pitch (note that auditory sensitivity to frequency ranges from ~20 and even up to ~20000 Hz, depending of the age and hearing of the person).

The second dimension is *loudness* that is related to the sound intensity so that sounds can be perceived as more or less loud as a function of the physical intensity (although in some case loudness and intensity may be partially independent).

A third and more complex dimension is *timbre* (related to the sound's spectrum composition and its unfolding in time) that enables us to distinguish and recognize the different voices of a mixed sound scene (e.g. a guitar from a piano) even when they are perceived as having the same pitch and loudness.

Sounds have also a *subjective duration* that is, strictly speaking, not a sensation exclusively related to sound but shared by all our senses.

In general terms, sound sensations are categorized in two: tones and noises. Tones give a clear sensation of pitch (e.g., the human voice, the sound of a music instrument, the chirping of birds) and can be concatenated in salient pitch patterns (e.g. melodies). Noises do not give a strong sensation of pitch (e.g. the fan of the air conditioning) and can hardly produce salient pitch patterns (e.g. melodies [12]). Although sounds can be described in terms of pitch, loudness, etc. we often describe them referring to the event that generated the sound and the types of materials involved in the sound source (e.g., "hammer on an anvil", "sound of a waterdrop")[13].

## 2.2 Extension to sound experience

Daily experiences with sounds can be categorized as perceptual, cognitive and emotional experiences[14,15].

In a *perceptual* experience, psychoacoustics plays an important role in determining how pleasant a sound is: The sharper, louder, rougher, and noisier a sound is the more unpleasant it will be perceived. Temporal aspects of the sound (i.e., duration, repetitiveness) also give rise to event perception and its evolution (e.g., car approaching). Furthermore, sound provides cues regarding the physical quality of its source (i.e., material, size, geometry and direction)[16]

In a *cognitive* experience, listeners are able to semantically distinguish their experiences with sounds and categorise them in terms of information regarding the sound event (i.e., source, action, location) and its conceptual associations (e.g., adventurous, playful).

In an *emotional* experience, the benefit/harm of the identified sound event to the task at hand is assessed and a sound is appraised whether it signals a potential threat (e.g., fire alarm) or poses opportunities for action (e.g., recognising the bike bell to move aside).

However, not all sound experiences can fit in discrete categories and all experiences with sound are contextual[17, 18]. While we may experience a sound perceptually unpleasant (e.g., a sharp and loud sound of an espresso machine), the context can turn this perceptually unpleasant experience into a functionally acceptable sound (i.e., all espresso machines produce such sounds as a result of their mechanical construction) or even a desirable one by means of the circumstantial associations (e.g., coffee machines endorsed by famous people) or cultural connotations (e.g., pleasure of drinking coffee in Portugal).

## 2.3 <u>Analysis</u>: sound experiences from existing tools

Overall, when designing sounding objects, the sound creation process can borrow knowledge from the object perception literature that analyses objects on *feature*, *object* and *scene* levels[19,20]. Thus, designers can address the featural aspects of sound in order to give form to the sound (e.g., an incrementally louder and repetitive sound can be perceptually salient by capturing attention, can indicate the evolution of an event, and be perceived as alarming or thrilling). However, designing these physical sound features should give rise to a meaningful whole that can solely be identified as a sound object (e.g., audible notification as alarms or approaching footsteps) and the sound object matches the scene it emerges in (e.g., medical care or game world). A coherence between the three dimensions of object perception[21] will pose less perceptual/cognitive load on the user, as the sound and its fittingness to the designated function or to its environment will be ensured and sound's capacity to fulfill a user's need will be achieved.

As far as sonification of objects is concerned, all these sensations can be exploited to communicate and represent quantities (as in data representation) and concepts (as messages to be conveyed such as size, shape, materials). In doing this we map one domain (the auditory sensation) onto another domain (e.g., luminances, sizes, etc.). Some associations are more natural than others because these associations can be frequently observed in nature. For example, in nature, low pitch sounds are usually associated with large objects whereas high pitch sounds are usually associated with small objects[22]. Alternatively, the mapping can be arbitrary and needs to be learnt: we can map pitch with distance (e.g., high pitch with large distance) although in nature such a relation doesn't exist. Pitch is perhaps the most investigated and mapped sensation. The reason for this, is that humans are very sensitive to pitch variations (an ability that we can improve with practice) and remember pitch relationships very well. For example, we can remember a sequence of pitches (i.e., a melody) after the first time we listen to it, whereas we may forget immediately a sequence of loudnesses[23]. In addition, pitch (but also loudness) has often a spatial connotation: we refer to pitch using the adjectives "high" and "low" and this is done in the same way in several music cultures[24].

Astronomers could use sound experience as an approach when they want to create pleasant, meaningful and contextual experiences when sonifying astronomical data (e.g., temperature fluctuations in sun observations) and space objects (e.g., milkyway, planets, or galaxies as a whole) or conveying a high-level concept (e.g., the stark beauty of a supernova). However, object identification notion will help congruent acoustic mappings of data-to-sound and better representations of space objects that fits a designated function, a space mission, or the research agendas.

## 3. From Sound Design to Sonic Information Design
### 3.1 Basics of sound design and sonification

Designing sounds means "to make an intention audible"[25]. A designed sound is new and constructed, and it represents something other than the sound itself. This can be an object, a concept, a dataset or a system. There are two intentions that need to be audible: form, which relates to sound quality, and function, which relates to what the sound communicates.

In sound design in general, the information portrayed needs to be clearly heard and correctly interpreted for the design to be considered successful. The history of sound design can be traced from Greek and Roman theatre to the development of new audio technology and media (radio, television, cinema, games, virtual reality) in the 20th century which generated a great variety of new methods for designing sounds. Recent research taps into this, and related knowledge and creative practice, to inform new methods for functional sound design, such as sonic interaction design and sonification[26,27,28,29].

The invention of the Geiger counter, at the beginning of the 20th century, is a well-known early example of sonification. The further development of electronics, computers and digital technology motivated the need for new ways to display and access information.

Sonification is a type of auditory display and sound design that is defined as aiming to "transform data relations into perceived relations in an acoustic signal for the purposes of facilitating communication or interpretation"[30]. The goals include increased accessibility, monitoring of dynamic processes, data mining, as well as the creation of new artistic experiences for audiences. Applications can be found in assistive technologies[31], health and environmental science[32,33,34,35,36], automotive engineering[37], mobile computing[38,39], intelligent alarms[40,41], technology-enhanced learning[42], and many more fields.

Auditory display techniques have been categorized as earcons (musical motifs)[43], auditory icons (everyday sounds)[44], audification (playback of data series at audio rates)[45], parameter mapping (mapping data parameters to sound parameters such as loudness or pitch)[46], and model-based sonification (map the dataset to a digital model that can be excited to make sounds)[47]. More recent methods include acoustic sonification (mapping the dataset to a 3D printed object that makes sounds acoustically)[48], and stream-based sonification (figure/ground gestalts that form auditory scenes)[49]. *Acoustic Sonification* may be of particular interest to astronomers with visual impairment because the physical object made from the data can be picked up, felt and explored by hand[50]. *Stream-based Sonification* uses psychoacoustic principles to group and segregate data mapped into auditory scenes[9] that are more like the sounds of the everyday world – thus, being perceived and understood through experiences of everyday listening[51] –, and that do not rely on musical training to hear, analyse and comprehend[52].

## 3.2 Creation: innovative tools for astronomical data sonification

The astronomy tools described in the first section use the parameter mapping technique where data values are played as notes on a musical instrument. Computer music software makes it easy to apply this technique using score-based interfaces. However, although this is a quick and simple technique, short term auditory memory is only about 2-4 seconds long[53] which makes it difficult to answer questions about longer sequences of notes.

SonoUno also uses the spearcon technique to support navigation of the user interface by Blind-Visually Impaired (BVI) astronomers. *Spearcons* use sped up speech to represent menu items and other parts of the interface, similar to the way screen readers can also use sped up text to speech. There is potential for *Earcons* and *Auditory Icons* to also be used to support auditory interfaces for BVI astronomers. Other techniques that have been explored by the data sonification community may be applicable and useful in astronomy. *Audification*, as used by Alexander[54] for the Heliosphere, was employed to play the data at audio rates so that the human ear did the spectral processing, rather than the computer – he found that human subjects were able to hear spectral features that they could not detect in graphic visualisations. *Fourier Transform*, as used by Sturm[55,56], was employed to sonify spectral data from Ocean Buoys as short sounds with different timbres. *Spectral Audification*, as used by Newbold[57], is a similar technique for the spectral analysis of chemicals.

These examples demonstrate the potential for astronomers to audify spectral data as timbres rather than note sequences, which may be more perceptually direct. *Model-based sonification*[58] is another interesting technique that could be applied in astronomy. Thomas Hermann describes one example, called a data sonogram, where the data points are mapped onto a simulated mass-spring system, so for example stars in an image could be nodes in such a system. The user initiates a shock wave that propagates spherically through the spring network which vibrates to produce the sound of that configuration of stars. An *Acoustic Sonification* could be made by mapping spectral values onto the parameters of a 3D shape that is 3D printed in a resonant metal so that it vibrates acoustically. Datasets could be held and felt, and differences between entire datasets could be heard immediately by tapping or scraping them. Many astronomical datasets have a lot of noisy background. *Stream-based* Sonification techniques could be used to design auditory scenes where where fast transients and weak signals perceptually emerge as auditory figures from the noisy background.

## 3.3 Extension to Sonic Information Design

The design of a sonification to provide useful information requires more than the arbitrary selection of a technique for mapping data into sound. *Sonic Information Design* is a user-centred method that requires consideration of issues such as the type of data, user task, information requirements, and audio display[59]. A designerly approach to sonification that includes stages of ideation, rapid prototyping and evaluation has been developed further[60]. The Data Sonification

Canvas provides a design-oriented approach that includes consideration of Users, Goals, Context, Functionality, Ways of Listening and Type of Sound[61,2].

## 4. From Psychoacoustics to Sonification Evaluation

### 4.1 Basics of psychoacoustics and experimental psychology

Psychoacoustics is the discipline that studies the relationships between a sound parameter (e.g., sound level) and an associated auditory sensation (e.g., loudness) obtained by measurement with human participants. In the present sub-section, methods are briefly presented. In the next sub-section, questions related to astronomical data sonification that can be approached by these methods are indicated.

Traditional psychoacoustical methods are unidimensional, and can be implemented either via direct or indirect methods.

Indirect methods are based on the measurement of thresholds (see method 1, in Table 1). Absolute threshold is the minimum detectable intensity of the sensation, whereas differential threshold is the smallest change in a sound to produce a Just-Noticeable Difference (JND). For instance, the JNDs for fundamental frequency (related to pitch) and spectral centroid (related to brightness) are of 0.8% and 4% respectively, for musicians, and 1.9% and 5% respectively[62], for non-musicians. The four usual methods to measure thresholds are the methods of constant stimuli, limits, adjustment, and the adaptive method[63].

Direct scaling methods (see method 2) rely on the ability of participants to assign a number proportional to their sensation, and at the end, the obtained relation expresses a direct sensation ratio in relation to the physical parameter. For instance, for a 1-kHz tone, a 10-dB increase leads to a doubling of the loudness[64].

For complex or real sounds, exploratory methods are usually adopted. They can be implemented in three main paradigms:

Dissimilarity ratings (see method 3) have been frequently adopted to investigate timbre of musical sounds[65,66] and environmental sounds[67]. Judgements are based on dissimilarity ratings among pairs of sounds, which are then represented by distances in a low-dimensional space using a Multidimensional scaling (MDS) technique.

Semantic scales (see method 4) are frequently used to assess auditory attributes (loudness, roughness, etc.) but also different psychological aspects of sounds such as appraisal judgments (e.g., preference). Judgements are based on direct evaluations on a k-point scale (k being odd and usually between 3 and 9) defined by a label (e.g., "dull - bright"). It is crucial that the participants correctly understand the meaning of the labels. To overcome any misunderstanding, a sound lexicon ('words4sounds') has been recently developed into the SpeaK environment (https://speak.ircam.fr/en/).

Sorting and identification tasks (see methods 5&6) are very commonly used in cognitive psychology to address the questions of identification and categorization of sound sources. Listeners are required to sort a corpus of sounds and to group them into as many classes as they want, or into a limited number of classes associated with labels in the case of an identification task. The resulting data are usually formatted in hierarchical structures (dendrogram) that represent clusters of sounds. Sorting tasks have been largely used to study the categorization of everyday soundscapes[68].

Finally, series of tests and analyses have been developed in the field of sound quality[69] to determine preference scales (see method 7). In addition, it should be emphasized that the classical psychoacoustical methods have been broadly used to study the perception of short and stationary sounds, however the fact a sound is time-based means the sonification is often more effective when the user is tightly embedded within a real-time evaluation. In this case, methods of continuous judgments can be considered (see method 8).

### 4.2 <u>Validation</u>: perceptual evaluation of astronomical data sonification tools

Most of the methods presented in the previous section are precisely detailed in [70]. In this section, a list of specific questions related to the sonification of astronomical data is posed. These questions could be used as reference and starting point to conceive the perceptual evaluation of a specific astronomy-focused sonification tool.

*Table 1. Basic evaluation methods, derived from psychoacoustics paradigms, described in Section 4.1. Each of the 8 methods belong to a certain methodological category: indirect or direct measurements (resp., 1, 2), dissimilarity, semantic, sorting or identification tasks (resp., 3, 4, 5, 6), preference judgements (7) and – shared with some of the previous ones – continuous evaluation along time (8). For each of these methods, the table contains examples of questions that could be potentially addressed within a validation protocol dedicated to astronomical data sonification tools*

|   | Method | Question that can be answered |
|---|---|---|
| 1 | Threshold measurements | Can the user perceive differences between characteristics of different astronomical objects?<br><br>**For example:** *Does the discrimination threshold on the auditory dimension used for the sonification account for the perceived change of a star brightness?* |
| 2 | Scaling methods | How should one auditory dimension vary for fitting the characteristics of an astronomical object?<br><br>**For example:** *Does the ratio in the auditory dimension correspond to the ratio of the depth of a transit in the light curve (relates to the size of the object relative to the host star)?* |
| 3 | Dissimilarity ratings | What are the main differences between multidimensional sonified astronomical objects?<br><br>**For example:** *If several characteristics of stars or galaxies are sonified by sounds made of several parameters (loudness, pitch, roughness, attack time, …), what are the most salient dimensions for the comparison? Are they weighted similarly?* |

| 4 | Semantic scales | What are the auditory profiles related to different words associated with different astronomical objects? |
| --- | --- | --- |
| | | **For example:** *Stars could be evaluated and compared in terms of sonic profiles; star #1 sounds bright, rough and <u>continuous</u>, although star #2 sounds also bright and rough, but <u>discontinuous</u>, which means unstable because of its internal structure.* |
| 5 | Sorting tasks | What is the most typical auditory configuration for a class of astronomical objects? |
| | | **For example:** *What are the different and similar shapes of light curves between several transits?* |
| 6 | Identification tasks | What are the sonic configurations that make it possible to classify different types of astronomical objects? |
| | | **For example:** *What is the boundary between two sound configurations that makes it possible to identify two different chemical fingerprints (related to the presence or absence of certain frequencies)? Does the boundary between two sound configurations allow to distinguish between a U and V shape in a transit light curve?* |
| 7 | Preference scales | Which is the preferred sound model for the sonification of a specific astronomical object? |
| | | **For example:** *What is the preference between astronomical data played slowly or quickly? What is the most pleasant/efficient among different sonifications, as for example tones vs pulses to explore a 2D spectroscopy?* |
| 8 | Continuous evaluations | Do users detect real-time changes in the sonification of relative position of astronomical moving objects? |
| | | **For example:** *Is it possible to detect changes in the intensity of light emitted by a galaxy by real-time continuous evaluation of its sonification?* |

## Conclusion

Finally, the interdisciplinary dialogue originally envisaged as a keystone of the Audible Universe approach started to be established in a rather concrete and fruitful way. A common 'playground', in which both communities – astronomers and sound scientists – brought their respective expertise and know-how, was basically delimited as the frame of a collaborative sound design process.

In this regard, from what inspired them in relation to existing tools, sound experts gave multiple insights in order to possibly create different design propositions – among which some of them, like acoustic sonification, could specifically be well adapted to inclusivity – and recommend relevant evaluation paradigms in terms of astronomical data sonification. Actually, basic knowledge in sound perception and cognition, sound design and sonification, and at last psychoacoustics and perceptual evaluation has started to bring new concepts and methods in the astronomy field, and furthermore to open new perspectives on how to observe, analyse, represent or transmit astronomical data.

But, as very often in the scientific domain, the Audible Universe workshop and its general approach have opened up more questions than they have practically resolved. For sure, many other forums and meetings will be required to address some issues raised during the first discussions (universality, standardization, multimodality, etc.) but also those that have not been explicitly formulated but are nonetheless of high importance, such as the role of emotions, the attention to multiculturality or the reflection on artificial intelligence … to be continued!

# Acknowledgements

We are grateful to the Lorentz Center for supporting the organization of the Audible Universe workshop in September 2021 and to the workshop participants for valuable and insightful discussions

# Author contributions

N.M. led the initiation, structuring and editing of this Perspective, the management of co-authors' contributions and the writing of Context of this Perspective, Introduction and Conclusion. E.Ö. and M.G. led the writing of From sound perception to sound experience, S.P. and S.B. led the writing of From sound design to sonic information design and R.B. and P.S. led the writing of From psychoacoustics to sonification evaluation. All co-authors participated in discussions about the content, and provided comments on the initial manuscript and feedback for the revised versions.

# Competing interests

The authors declare no competing interests.

# End Matter

Corresponding author.

Name: Nicolas MISDARIIS
Address : Ircam STMS Lab (Ircam-CNRS-SU-MinCult)
            1 Place Igor Stravinsky, 75004 Paris, France.
Phone Nb.: +33 1 4478 1350.
E-mail address: nicolas.misdariis@ircam.fr